\documentclass[aps,prl,floatfix,epsfig,twocolumn,showpacs,preprintnumbers]{revtex4}
\usepackage{eurosym}
\usepackage{amssymb}
\usepackage{amsmath}
\usepackage{graphicx}
\usepackage{epstopdf}
\usepackage{dcolumn}
\usepackage{bm}
\usepackage[pdfstartview=FitH, CJKbookmarks=true, bookmarksnumbered=true, bookmarksopen=true, pdfborderstyle={/S/U}, pdfborder={0 0 1}, colorlinks=true, linkcolor=blue, anchorcolor=green, citecolor=blue]{hyperref}

\DeclareMathAlphabet      {\mathbf}{OT1}{cmr}{bx}{n}
\begin{document}

\title{Calculations of Spin Fluctuation Spectral Functions $\alpha^{2}F$ in
High-Temperature Superconducting Cuprates}
\author{Griffin Heier and Sergey Y. Savrasov}
\affiliation{Department of Physics, University of California, Davis, CA 95616}
\date{\today }

\begin{abstract}
Spin fluctuations have been proposed as a key mechanism for mediating
superconductivity, particularly in high-temperature superconducting
cuprates, where conventional electron-phonon interactions alone cannot
account for the observed critical temperatures. Traditionally, their role
has been analyzed through tight--binding based model Hamiltonians. In this
work we present a method that combines density functional theory with a momentum- and
frequency-dependent pairing interaction derived from the Fluctuation
Exchange (FLEX) type Random Phase Approximation (FLEX-RPA) to compute
Eliashberg spectral functions $\alpha ^{2}F(\omega )$ which are central to
spin fluctuation theory of superconductivity. We apply our numerical
procedure to study a series of cuprates where our
extracted material specific $\alpha ^{2}F(\omega )$ are found to exhibit
remarkable similarities characterized by a sharp peak in the vicinity of
40--60 meV and their rapid decay at higher frequencies. Our exact
diagonalization of a linearized BCS gap equation extracts superconducting
energy gap functions for realistic Fermi surfaces of the cuprates and
predicts their symmetry to be $d_{x^{2}-y^{2}}$ in all studied systems. Via
a variation of on--site Coulomb repulsion $U$ for the copper $d$--electrons
we show that that the range of the experimental values of $T_{c}$ can be
reproduced in this approach but is extremely sensitive to the proximity of
the spin density wave instability. These data highlight challenges in
building first--principle theories of high temperature superconductivity but
offer new insights beyond previous treatments, such as the confirmation of
the usability of approximate BCS--like $T_{c}$ equations, together with the
evaluations of the material specific coupling constant $\lambda $ without
reliance on tight--binding approximations of their electronic structures.
\end{abstract}

\date{\today }
\maketitle

\section{I. Introduction}

Strong antiferromagnetic spin fluctuations have long been considered a
potential mechanism for unconventional superconductivity, following early
theoretical proposals\cite{Varma_1986,Scalapino_1986}. Shortly after these
foundational papers the first high--temperature superconductor was discovered%
\cite{SC_Discover}. In spin fluctuation theory, the interaction vertex is
commonly modeled using the Random Phase Approximation (RPA), which serves as
a simplified representation of the true interaction vertex. Despite its
simplicity, the RPA approach provides good qualitative results, particularly
in capturing the divergent behavior as the spin density wave (SDW) is
approached, an essential feature of many unconventional superconductors. The
RPA vertex also naturally takes into account Fermi surface nesting, a
phenomenon frequently observed in these systems\cite{Unconventional_review}.
As a result, spin fluctuations have remained a popular candidate for
explaining superconductivity and have been applied extensively in the
theoretical analysis of various unconventional superconductors. These
materials include cuprates\cite{cu_flex1,cu_flex2,cu_flex3,cu_flex4},
nickelates\cite{ni_flex1,ni_flex2}, ruthenates\cite{ru_flex1,ru_flex2},
cobaltates\cite{co_flex1}, ironates\cite{fe_flex1,fe_flex2,fe_flex3,fe_flex4}%
, and heavy fermion\cite{hf_flex1,hf_flex2} systems.

At a given level of doping, theoretical methods typically aim to predict the
onset of superconductivity, characterized by the transition temperature $%
T_{c}$, as well as the symmetry of the resulting superconducting gap
function. Most approaches utilize simple tight--binding models with a
limited number of orbitals and band structures, found using the Local
Density Approximation\cite{LDA_review} (LDA), which serve as an input to the
Hubbard model Hamiltonian. The latter is subsequently solved by an available
many--body technique, such, for example, as the Fluctuational Exchange
Approximation (FLEX) \cite{FLEX}. FLEX is a diagrammatic method that
includes RPA diagrams given by the particle--hole ladders and bubbles as
well as particle--particle ladder diagrams. It is however known that in the
vicinity of the SDW instability the most divergent terms are due to the
particle--hole ladders\cite{Paramagnons,BerkSchrieffer}, and the
contribution from the\ particle--particle ladders can be neglected\cite%
{Muller}.

Although a lot of past studies of spin fluctuational superconductivity
utilized RPA and FLEX\cite{FLEX-Review}, the extension of this technique has
been proposed recently by combining them with dynamical mean field theory
(DMFT) \cite{FLEX+DMFT}. It has resulted in reproducing doping dependence of
the $T_{c}$ in the simulation of the two--dimensional Hubbard model on the
square lattice relevant to the cuprates. More accurate Quantum Monte Carlo
simulations provide further ways to improve this approach\cite{Maier,Maier2}%
, and realistic electronic structures can be incorporated via the use of
theories such as LDA+U\cite{LDA+U} and LDA+DMFT\cite{LDA+DMFT} by utilizing
the method of projectors.

Our own recent combination of the LDA with FLEX and RPA\cite{LDA+FLEX}\
(LDA+FLEX) has allowed us to extend these methodologies to incorporate $%
\mathbf{k}$-- and $\omega $ dependence of the electronic self--energy by
evaluating dynamical spin and charge susceptibilities in a restricted
Hilbert space given by correlated orbitals. With a further extension of this
method to compute the FLEX interaction vertices for the Cooper pairs\cite%
{LDA_FLEX1}, the description of unconventional superconductors in a
realistic material framework became possible. Our most recent applications
to HgBa$_{2}$CuO$_{4}$\cite{LDA_FLEX1} recovered the known $d_{x^{2}-y^{2}}$
symmetry of the order parameter here, and predicted various competing
pairing states in a recently discovered superconducting nickelate La$_{3}$Ni$%
_{2}$O$_{7}$ \cite{LDA_FLEX2} and in vanadium kagome metal CsV$_{3}$Sb$_{5}$%
\cite{LDA_FLEX3}.

Estimates of the transition temperature $T_{c}$ represent an additional
challenge. In electron--phonon superconductors, it is typically calculated
using Bardeen-Cooper--Schrieffer (BCS) theory\cite{BCS_theory} where the
central quantity to determine is the coupling constant of the interaction $%
\lambda $, from which the $T_{c}$ can be found via the McMillan equation\cite%
{McMillan_1968}. Although this approach is straightforward, these estimates
exhibiting exponential sensitivity to the errors in $\lambda $ and
uncertainties in the treatment of the Coulomb pseudopotential $\mu ^{\ast }$
usually serve to obtain the correct order of magnitude for the $T_{c}$. More
rigorous Eliashberg theory of superconductivity requiring the input of the
spectral function $\alpha ^{2}F\left( \omega \right) $ and the evaluation of
the screened Coulomb interaction can be utilized to improve the accuracy of
these estimates\cite{Gross}, but in a semiquantitative way, the McMillan
theory is known to suffice in many cases.

There are good reasons for the successes of the BCS approach, in particular,
because in the electron--phonon superconductors the $\alpha ^{2}F\left(
\omega \right) $ is limited by the Debye frequency $\omega _{D}$ which is
small comparing to the typical energies of the electronic states. For
unconventional superconductors, technically, the Coulomb interaction vertex
as, for example, given by FLEX, does not provide such a small parameter,
therefore, many model calculations based on diagonalizing the BCS--type
equations in case of spin fluctuations\cite{FLEX-Review} necessitate
additional assumptions. To overcome this difficulty, the Eliashberg equation
can be solved directly on the imaginary Matsubara axis and predict the $%
T_{c} $ directly\cite{Arita2000,FLEX-Review}, but the methodology hides the
transparency of the BCS approach which has allowed to guide the discoveries
of many conventional superconductors by searching for the materials with
large $\lambda .$

Here, we provide new insights to the superconductivity of single-- and
double--layer cuprates\cite%
{Hg-Supra,Sr2CuO2,Tl-Supra1,BM,IL-Supra,123-Supra,Tl-Supra2,BSCO-Supra} by
calculating their spin fluctuational spectral functions $\alpha ^{2}F(\omega
)$ for a given pairing symmetry on the basis of frequency-- and
momentum--resolved FLEX--RPA pairing interaction and by utilizing their
realistic LDA based electronic energy bands and wave \ functions. We first
exactly diagonalize the linearized BCS gap equation and extract the highest
eigenstate and the eigenvector as a function of the Fermi momentum set on a
realistic Fermi surface of the material. We find the corresponding
superconducting gap functions to be of $d_{x^{2}-y^{2}}$ symmetry in all
studied cuprates for a wide range of on--site Coulomb repulsions $U$ and
dopings that we scan during our simulations. We further utilize the
Kramers--Kronig relations for the frequency dependence of the FLEX\ vertex,
and its averaging over the eigenvectors provides a frequency resolution for
each eigenstate. This gives rise to the definition of the spectral functions 
$\alpha ^{2}F(\omega )$ whose inverse frequency moments correspond to the
eigenvalues of the BCS\ gap equation.

Our calculated material specific $\alpha ^{2}F(\omega )$ are found to
exhibit remarkable similarities characterized by a sharp peak in the
vicinity of 40--60 meV which has been known to exist from a wealth of
experimental data in cuprates such, e.g., imaginary spin susceptibility
accessible via the neutron scattering experiment\cite{INS}, the existence of
a 40 meV resonance visible in the superconducting state\cite{40meV}, as well
as the kinks in the one--electron spectra seen at this energy range in angle
resolved photoemission (ARPES) experiment\cite{SF_3}. We also find that $%
\alpha ^{2}F(\omega )$ decays rapidly at higher frequencies which justifies
the use of the BCS--type treatment in the cuprates.

The maximum eigenvalue $\lambda _{\max \text{ }}$deduced from this procedure
represents a spin fluctuational coupling constant similar to the
electron--phonon (e--p) $\lambda _{e-p}.$ We find it to be particularly
large when the system is close to SDW, which in this theory appears for a
particular value of $U=U_{SDW},$ that corresponds to the divergent static
spin susceptibility at some wavevector. To allow evaluation of the\ $T_{c}$
via the McMillan theory, we calculate the normal state self--energy and
extract the electronic mass enhancement $m^{\ast }/m_{LDA}=1+\lambda _{sf}.$
The mass enhancement is also large in the vicinity of SDW$.$ However, the
evaluated renormalized coupling constant $\lambda _{eff}=\lambda _{\max 
\text{ }}/(1+\lambda _{sf})$ that determines the BCS\ $T_{c}$ was found to
be modest unless we tune $U$ to be within a few percent of $U_{SDW}$. \
Several numerical approaches to find $U$ were proposed and applied to the
cuprates in the past, such as constrained Random Phase Approximation (cRPA)%
\cite{cRPA_1,cRPA_2}, or fitting $U$ to reproduce some experiments, but such
a great sensitivity to the proximity to $U_{SDW}$ rules out the
applicability of these methods here.

As the oldest and most extensively studied family of unconventional
superconductors, cuprates provide a wealth of published mass enhancement
data that can be extracted from ARPES, quantum oscillation and specific heat
experiments. This data offer additional comparisons with the values of $%
\lambda _{sf}$ that we calculate in the hope to find further constrains for
the material specific $U$ that is needed to predict the $T_{c}$. However,
when comparing experimental data with theoretical predictions, different
experimental techniques require specific considerations. For example, ARPES
measures a two--dimensional spectral function, from which the Fermi velocity 
$v_{F}^{\ast }$ can be extracted. Quantum oscillation studies use magnetic
fields to induce oscillations around the Fermi surface. The frequency of
these oscillations provides a measurement of the effective mass $m^{\ast }$.
Specific heat measurements give the electronic specific heat coefficient $%
\gamma ^{\ast }$ found from all bands. Many of these experiments are
surveyed in our work but the spread in the deduced values of $\lambda _{sf},$
while being well within our theoretical predictions, does not offer further
tuning of the constrains on the best determination of $U$.

The results presented here highlight challenges in building first--principle
theories of high--temperature superconductivity but still offer new insights
that go beyond previous treatments, such as the calculations of $\alpha
^{2}F $'s$,$ the confirmation of the usability of approximate BCS--like $%
T_{c}$ equations, and the evaluation of the material specific coupling
constant $\lambda $ without reliance on the tight--binding approximations of
their electronic structures.

This paper is organized as follows. In Section II\ we survey our LDA+FLEX\
method and describe the numerical procedure to calculate the spin
fluctuational spectral functions $\alpha ^{2}F(\omega ).$ In Section III we
report our results on the calculated behavior of superconducting energy
gaps, $\alpha ^{2}F(\omega ),$ the coupling constants $\lambda _{\max }$ and
the mass enhancement parameters $\lambda _{sf}$ for a series of single-- and
double--layer high--$T_{c}$ cuprates. In Section IV we analyze available
experimental data on the electronic mass renormalization deduced from ARPES%
\cite%
{Vishik_2014,Birgeneau_1994,Peets_2007,ARPES-MASS2,Uchida_2003,Raffy_2005,Follath_2006,Muller_1988,Vishik_2009,Bi_overdoped,Berger_2006,Shen_1999}%
, quantum oscillations\cite%
{Harrison_2019,Greven_2013,Greven_2016,Carrington_2010,Wilson_2009,Crooker_2021,Proust_2008,Taillefer_2007,Klein_2018}
and measured values of the specific heat\cite%
{Klein_2020,Tallon_1994,Hussey_2003,Klein_2019}. Section V is the conclusion.

\section{II. Method}

The LDA+FLEX formalism describes a spin--dependent interaction between
electrons\cite{LDA_FLEX1,LDA_FLEX2}, with the common approximation that
interactions occur near the Fermi surface. The interaction is denoted by

\begin{equation}
K^{\nu _{1}\nu _{2}\nu _{3}\nu _{4}}(\mathbf{r}_{1},\mathbf{r}_{2},\mathbf{r}%
_{3},\mathbf{r}_{4},\omega ).  \label{INT}
\end{equation}%
In the non--relativistic formalism, the spin space remains fully invariant.
This vertex can be decomposed into charge $K^{c}$ and spin $K^{s}$
components based around the Pauli matrices%
\begin{equation}
K^{\nu _{1}\nu _{2}\nu _{3}\nu _{4}}=\frac{1}{2}\delta _{\nu _{1}\nu
_{3}}\delta _{\nu _{2}\nu _{4}}K^{c}-\frac{1}{2}\mathbf{\sigma }_{\nu
_{1}\nu _{3}}\mathbf{\sigma }_{\nu _{2}\nu _{4}}K^{s}.
\end{equation}%
This decomposition facilitates a transformation into singlet--triplet space,
where

\begin{equation}
K^{(S)}=\frac{1}{2}K^{c}-\frac{1}{2}E_{s}K^{s}
\end{equation}%
with $E_{S=0}=-3,E_{S=1}=1$. Projecting the scattering from the Cooper pair
wave functions $\Psi _{\mathbf{k}j,SS_{z}}(\mathbf{r}_{1},\mathbf{r}_{2})$
onto $K$ further reduces the vertex. Here $\mathbf{k}$ is the momentum
vector, $j$ the band index, $S$ and $S_{z}$ are the spin and its
z--projection for the Cooper pair. This results in the matrix $M$:

\begin{equation}
M_{\mathbf{k}j\mathbf{k}^{\prime }j^{\prime }}^{(S)}(\omega )=\langle \Psi _{%
\mathbf{k}j,SS_{z}}|K^{(S)}(\omega )|\Psi _{\mathbf{k}^{\prime }j^{\prime
},SS_{z}}\rangle  \label{MAT}
\end{equation}%
Due to the Bloch--wave property of the Cooper pair wave functions$,$ the
lattice Fourier transforms with $\mathbf{k}$ and $\mathbf{k}^{\prime }$ of
the interaction

\begin{eqnarray}
&&K_{\mathbf{k,k}^{\prime }}^{(S)}(\mathbf{r}_{1},\mathbf{r}_{2},\mathbf{r}%
_{3},\mathbf{r}_{4},\omega )=\sum_{R_{1,2,3,4}}e^{-i\mathbf{k}(\mathbf{R}%
_{1}-\mathbf{R}_{2})}e^{i\mathbf{k}^{\prime }(\mathbf{R}_{3}-\mathbf{R}_{4})}
\notag \\
&&K^{(S)}(\mathbf{r}_{1}\mathbf{-R}_{1},\mathbf{r}_{2}\mathbf{-R}_{2},%
\mathbf{r}_{3}-\mathbf{R}_{3},\mathbf{r}_{4}-\mathbf{R}_{4},\omega ),
\end{eqnarray}%
are only relevant in the expression (\ref{MAT}), where one lattice sum
should be dropped out due to translational periodicity.

The Cooper pair wave functions are constructed from single--electron states,
easily accessible in density functional theory (DFT). The formidable problem
is the evaluation of the pairing interaction $K^{(S)}$. This vertex is
approximated to operate for a correlated subset of electrons that are
introduced with help of site dependent projector operators $\phi _{a}(%
\boldsymbol{r})=\phi _{l}(r)i^{l}Y_{lm}(\hat{r})$. These operators originate
from the one--electron Schr\"{o}dinger equation taken with a spherically
symmetric part of the full potential. The Hilbert space inside the
correlated site reduces the full orbital set to a subset of correlated
orbitals, such as the $l=2$ orbitals for Cu. Decomposing the $K^{(S)}$
vertex in this manner yields the representation

\begin{eqnarray}
&&K_{\mathbf{k,k}^{\prime }}^{(S)}(\mathbf{r}_{1},\mathbf{r}_{2},\mathbf{r}%
_{3},\mathbf{r}_{4},\omega
)=\sum_{a_{1}a_{2}a_{3}a_{4}}K_{a_{1}a_{2}a_{3}a_{4}}^{(S)}(\mathbf{k},%
\mathbf{k}^{\prime },\omega )\times  \notag \\
&&\phi _{a_{1}}(\boldsymbol{r}_{1})\phi _{a_{2}}(\boldsymbol{r}_{2})\phi
_{a_{3}}^{\ast }(\boldsymbol{r}_{3})\phi _{a_{4}}^{\ast }(\boldsymbol{r}%
_{4}).
\end{eqnarray}

Evaluating this vertex with FLEX, using the on--site Coulomb interaction
matrix $I_{a_{1}a_{2}a_{3}a_{4}}$ gives the final vertex equation:

\begin{equation}
\hat{K}=\hat{I}+\hat{I}\left( \hat{\chi}-\frac{1}{2}\hat{\pi}\right) \hat{I},
\label{CHI}
\end{equation}%
where the interacting susceptibility $\hat{\chi}=\hat{\pi}\left( \hat{1}-%
\hat{I}\hat{\pi}\right) ^{-1}$ contains the non--interacting polarizability $%
\hat{\pi}$, and the subtraction of $\frac{1}{2}\hat{\pi}$ removes the single
bubble diagram that appears in both the bubble and ladder series. The $\hat{I%
}$ matrix is built off the Hubbard term $U$ and is local in space. This
locality condition allows $K_{a_{1}a_{2}a_{3}a_{4}}^{(S)}(\mathbf{k},\mathbf{%
k}^{\prime },\omega )$ to be only dependent on $\mathbf{k}\pm \mathbf{k}%
^{\prime }$. Taking the full interactions is computationally demanding,
scaling as $N_{atom}^{4}N_{orb}^{4}$. However, as previously reviewed\cite%
{LDA_FLEX1}, the local approximation reduces the complexity to scale with $%
N_{atom}^{2}$, and the use of only correlated orbitals minimizes the $%
N_{orb}^{4}$ term, making the computations feasible.

The resulting matrix elements of the pairing interaction taken in Eq.(\ref%
{MAT}) for $\omega =0$ are Hermitian and are used in the linearized BCS gap
equation

\begin{equation}
\Delta _{S}(\mathbf{k}j)=-\sum_{\boldsymbol{k}^{\prime }j^{\prime }}M_{%
\mathbf{k}j\mathbf{k}^{\prime }j^{\prime },{Re}}^{(S)}(0)\frac{\tanh
(\epsilon _{\mathbf{k}^{\prime }j^{\prime }}/2T_{c})}{2\epsilon _{\mathbf{k}%
^{\prime }j^{\prime }}}\Delta _{S}(\mathbf{k}^{\prime }j^{\prime }).
\end{equation}%
Here $\epsilon _{\mathbf{k}j}$ denotes the non--interacting band structure
from which the superconducting state evolves, and $T_{c}$ is the critical
temperature, below which superconductivity exists. We focus only on
interactions occurring near the Fermi surface, taking the BCS approximation
that all other interactions beyond $\omega _{c}$ are zero. As a standard
practice, we also multiply the left part of this equation with a parameter $%
\varepsilon $ playing the role of the eigenvalue, and treat $\Delta _{S}(%
\mathbf{k}j)$ as the eigenvector. This results in 
\begin{eqnarray}
-\ln \left( \frac{1.134\omega _{c}}{T_{c}}\right) &&\sum_{j^{\prime
}}\int_{FS}\frac{dS_{\mathbf{k}^{\prime }}}{|v_{\mathbf{k}^{\prime
}j^{\prime }}|}M_{\mathbf{k}j\mathbf{k}^{\prime }j^{\prime },{Re}%
}^{(S)}(0)\Delta _{S}(\mathbf{k}^{\prime }j^{\prime })  \notag \\
&=&\varepsilon \Delta _{S}(\mathbf{k}j),  \label{BCS}
\end{eqnarray}%
where the integration is extended over the Fermi surface only. To view this
expression as diagonalization in $\mathbf{k}j\mathbf{k}^{\prime }j^{\prime }$
indexes, we introduce the renormalized eigenvalues $\lambda =$ $\varepsilon
/\ln \left( \frac{1.134\omega _{c}}{T_{c}}\right) $, and, since the physical
solution for $\Delta _{S}(\mathbf{k}j)$ is given when the highest eigenvalue 
$\varepsilon $ becomes unity, the highest renormalized eigenvalue $\lambda
_{\max }$ is the coupling constant that produces the famous BCS equation for 
$T_{c}=1.134\omega _{c}\exp (-1/\lambda _{\max }).$

McMillan's $T_{c}$ equation uses this coupling constant to determine $T_{c}$%
, but unlike the BCS, it also takes renormalization and higher--energy
damping terms into account. The coupling constant $\lambda _{\mathrm{max}}$,
the mass enhancement factor $\lambda _{sf}+1$ and the higher energy Coulomb
repulsion term $\mu _{\mathrm{m}}^{\ast }$ make up the 3 parameters in the
effective coupling constant used in the $T_{c}$ equation, analogous to
McMillan's:%
\begin{equation}
T_{c}\approx \omega _{c}\exp \left( -1/\lambda _{eff}\right)  \label{TC}
\end{equation}

\begin{equation}
\lambda _{eff}=\frac{\lambda _{\mathrm{max}}-\mu _{\mathrm{m}}^{\ast }}{%
\lambda _{sf}+1}  \label{LAMBDA}
\end{equation}

The cutoff frequency $\omega _{c}$ is often taken to be the resonant peak
seen in neutron scattering experiments\cite{SF_2,SF_review}, found to be
about 40meV. Similar energy scales are seen in photoemission experiments as
a kink in the emission spectrum\cite{SF_3}. The renormalization constant $%
\lambda _{sf}$ can be derived from the normal state self--energy at the
Fermi surface:

\begin{equation}
\lambda _{sf}=-\langle \frac{\partial \Sigma (\mathbf{k},\omega )}{\partial
\omega }|_{\omega =0}\rangle _{FS}.  \label{MASS}
\end{equation}%
The parameter $\mu _{\mathrm{m}}^{\ast }$ is usually small in standard $s$%
--wave superconductors (\symbol{126}0.1) but here\ it refers to the same
pairing symmetry as the eigenvalue $\lambda _{\mathrm{max}}.$\ We expect it
to be negligible for the $d$--wave symmetry as the projection of the
screened Hubbard interaction onto $d_{x^{2}-y^{2}}$ cubic harmonic is known
to be very small \cite{Alexandrov}.

Finally, we describe the calculation of the Eliashberg spectral function $%
\alpha ^{2}F(\omega ).$ Since the real and imaginary parts of the
interaction (\ref{INT}) obey the Kramers--Kronig relation, we can relate the
matrix elements (\ref{MAT}) that appear for $\omega =0$ in the solution of
the BCS equation (\ref{BCS}) to the matrix elements taken over the imaginary
part of the interaction (\ref{INT}):

\begin{equation}
M_{\mathbf{k}j\mathbf{k}^{\prime }j^{\prime },{Re}}^{(S)}(0)=\frac{2}{\pi }%
\int_{0}^{\infty }\frac{M_{\mathbf{k}j\mathbf{k}^{\prime }j^{\prime },{Im}%
}^{(S)}(\omega )d\omega }{\omega }.
\end{equation}%
Let us represent the eigenvalues of (\ref{BCS}) as the averages over the
eigenvectors%
\begin{eqnarray}
&&\lambda _{\max }\equiv \lambda _{\mathrm{Re}}(0)=\sum_{jj^{\prime }}\int
\int_{FS}\frac{dS_{\mathbf{k}}}{|v_{\mathbf{k}j}|}\frac{dS_{\mathbf{k}%
^{\prime }}}{|v_{\mathbf{k}^{\prime }j^{\prime }}|}\times  \notag \\
&&\Delta _{S}^{\ast }(\mathbf{k}j)M_{\mathbf{k}j\mathbf{k}^{\prime
}j^{\prime },{Re}}^{(S)}(0)\Delta _{S}(\mathbf{k}^{\prime }j^{\prime }).
\end{eqnarray}%
We arrive to the spectral resolution for the eigenvalues%
\begin{equation}
\lambda _{\mathrm{Re}}(0)=\frac{2}{\pi }\int_{0+}^{\infty }\frac{\lambda _{%
\mathrm{Im}}(\omega )d\omega }{\omega },
\end{equation}

\begin{eqnarray}
&&\lambda _{\mathrm{Im}}(\omega )=\sum_{jj^{\prime }}\int \int_{FS}\frac{dS_{%
\mathbf{k}}}{|v_{\mathbf{k}j}|}\frac{dS_{\mathbf{k}^{\prime }}}{|v_{\mathbf{k%
}^{\prime }j^{\prime }}|}\times  \notag \\
&&\Delta _{S}^{\ast }(\mathbf{k}j)M_{\mathbf{k}j\mathbf{k}^{\prime
}j^{\prime },{Im}}^{(S)}(\omega )\Delta _{S}(\mathbf{k}^{\prime }j^{\prime
}).
\end{eqnarray}%
Thus, $\lambda _{\mathrm{Im}}(\omega )$ is related to the Eliashberg
spectral function for every eigenstate of the BCS equation as follows: 
\begin{equation}
\alpha ^{2}F(\omega )=\frac{1}{\pi }\lambda _{\mathrm{Im}}(\omega ).
\end{equation}

\section{III. Results}

The method described above allows for the complete calculation of material
specific superconducting energy gaps, $\alpha ^{2}F(\omega )$ and coupling
constants without the need for band structure approximations. It allows us
to resolve the gap functions and the corresponding pairing symmetries for
realistic Fermi surfaces which is a step beyond previous tight--binding
calculations.

We use the full potential linear muffin--tin orbital method \cite{FPLMTO} to
calculate LDA energy bands and wave functions for a series of single-- and
double layer cuprates. We then utilize the LDA+FLEX(RPA) evaluation of the
pairing interaction $K_{a_{1}a_{2}a_{3}a_{4}}^{(S)}(\mathbf{q,}\omega )$ on
three dimensional grids of the $\mathbf{q}$ points in the Brillouin Zone. \
The Fermi surface is triangularized onto small areas described by several
thousands of Fermi momenta for which the matrix elements of scattering
between the Cooper pairs, Eq.(\ref{MAT}), are evaluated. The\ linearized
BCS\ gap equation (\ref{BCS}) is then exactly diagonalized and the set of
eigenstates is obtained for both $S=0$ and $S=1$ pairings. The highest
eigenvalue $\lambda _{\max }$ represents the physical solution and its
eigenvector corresponds to the superconducting gap $\Delta _{S}(\mathbf{k}j)$%
.

We use Hubbard interaction parameter $U$ for the d--electrons of Cu as the
input to this simulation. We first vary $U$ to determine\ the divergency of
the static spin susceptibility which usually occurs at or very close to the
antiferromagnetic ordering wavevector $\mathbf{q}_{AF}=(\pi ,\pi ,0)$. This
sets the upper bound, $U_{SDW},$ whose value for each cuprate is listed in
Table 1. We can subsequently vary $U$ for the range of values less than $%
U_{SDW}$.

We also introduce the doping by holes using the virtual crystal
approximation. A whole range of dopings $\delta \leq 0.5$ (we refer $\delta $
per single CuO$_{2}$ plane) is studied in our work but, since within RPA the
SDW instability always exist for some $U_{SDW}$ at any doping$,$ the doping
dependence of the results cannot be traced in our method. We therefore
present most of our results for the doping level equal to 0.1 hole per CuO$%
_{2}$ plane unless otherwise noted.

\subsection{a. Pairing Symmetries}

{}The result of our simulation for superconducting energy gaps is shown on
Fig. 1 where the calculated $\Delta _{S=0}(\mathbf{k}j)$ exhibit a much
celebrated $d_{x^{2}-y^{2}}$-- symmetry for 8 cuprates that are studied in
this work. The blue/red color corresponds to negative/positive values of $%
\Delta .$ The zeroes of the gap function are along (11) direction which are
colored in grey. Fig. 1 utilizes the doping level equal to 0.1 hole per CuO$%
_{2}$ plane, although the result is robust for the range of dopings $\delta
\leq 0.3$ and also for the range of $U$<$U_{SDW}$ that we used in the
simulation. The effect of higher dopings $\delta =0.4-0.5$ has been checked,
but at those levels the gap function was found to develop a rather complex
sign--changing behavior along the lobes by acquiring higher order harmonics.
Such solutions would carry an additional kinetic energy and should be less
favorable energetically.

\begin{figure*}[tbp]
\includegraphics[height=0.498\textwidth,width=1.00\textwidth]{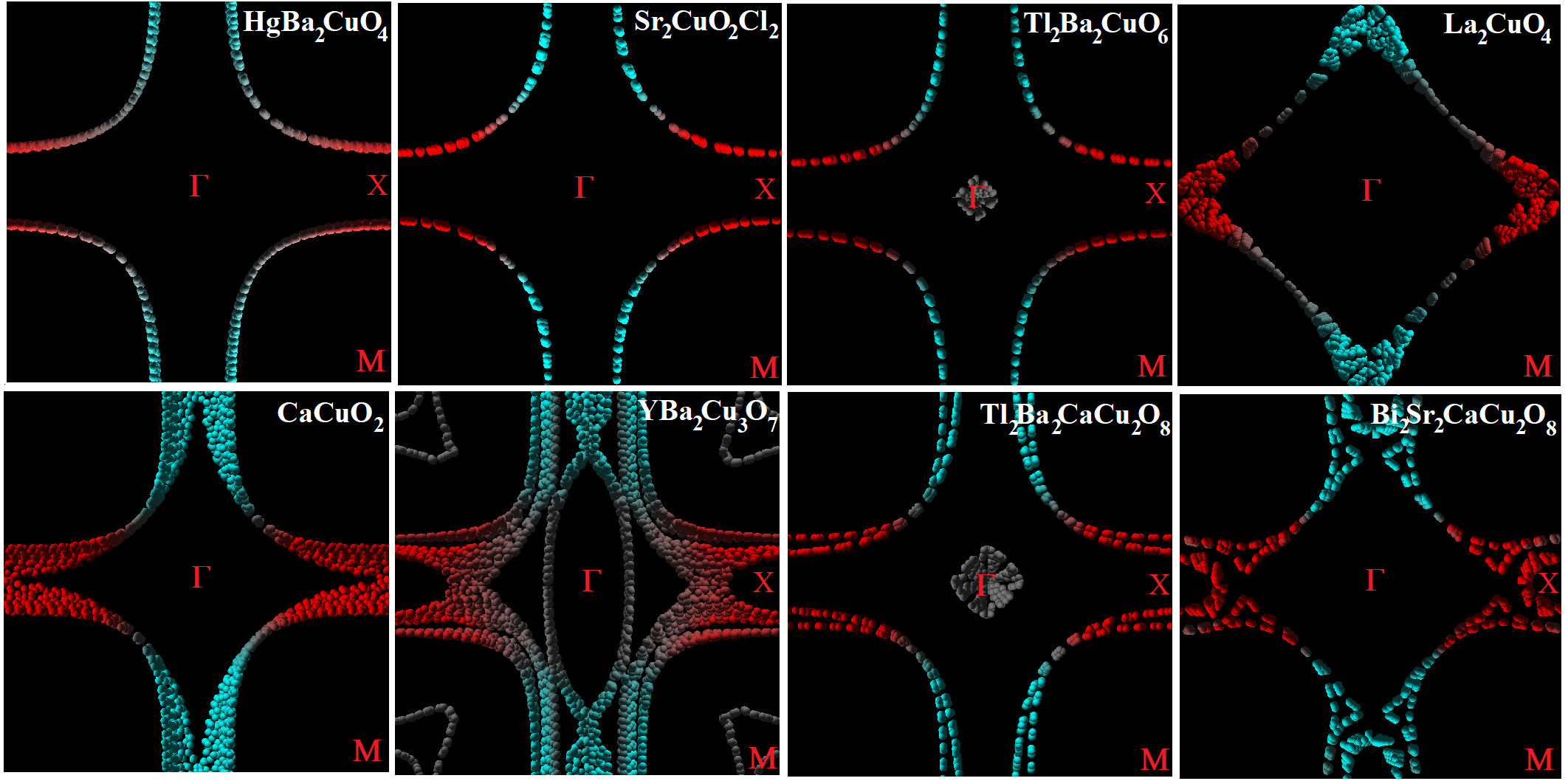}
\caption{{}Superconducting energy gap functions $\Delta (\mathbf{k})$ for 8
cuprates as a function of the Fermi surface wave vector (top view of the 3
dimensional Fermi surface is shown) calculated using numerical solution of
the linearized BCS gap equation with the singlet pairing interaction
evaluated via the LDA+FLEX(RPA)\ approach. The $d_{x^{2}-y^{2}}$ pairing
symmetry is seen in all cases as a sliding color scale with red
corresponding to a large positive gap, blue a large negative gap, and gray
the zero gap. The doping level is set to 0.1 holes per CuO$_{2}$ layer.}
\end{figure*}

We now turn to the material dependence of our results. $\mathrm{%
HgBa_{2}CuO_{4}}$ is one of the simplest single--layer cuprates, with large
spacer layers separating the CuO$_{2}$ superconducting planes. As a result,
its Fermi surface is made of just four hole pockets forming a kind of
rounded cross shape. This fundamental shape is present in all of the
cuprates, though the Fermi surface of $\mathrm{La_{2}CuO_{4}}$ is much more
diamond--like. Omitting the complications introduced by multiple bands, such
as those in $\mathrm{YBa_{2}Cu_{3}O_{7}}$, the addition of holes gradually
transforms the Fermi surface of each cuprate to resemble that of $\mathrm{%
La_{2}CuO_{4}}$, reducing the electron--filled area within the Brillouin
zone. This also alters the dominant wavevector $\mathbf{q}$ that leads to
the maximum Fermi surface nesting.

$\mathrm{YBa_{2}Cu_{3}O_{7}}$ is a particularly interesting material. The CuO%
$_{2}$ planes, as well as the CuO chains that occur underneath, lead to a
complicated Fermi surface. There are a number of pockets appearing from the
many band crossings that result from the Cu--O planes and chains, not all of
them superconducting. While red and blue represent positive and negative
signs for the superconducting gap function, gray represents values closer to
zero. In materials where the entirety of the Fermi surface contributes to
the interaction, the gray points represent nodes where the gap vanishes.
This can be seen in all of the superconducting plots in Fig. 1, where the
red section meets the blue section.

The $d_{x^{2}-y^{2}}$ Cu orbitals are known to be correlated and lead to
superconductivity, but this is not true of all orbitals, even those at the
Fermi surface. In those cuprates where uncorrelated states are present at
the Fermi level, the associated bands fail to develop a superconducting gap.
In particular, materials such as $\mathrm{YBa_{2}Cu_{3}O_{7}}$, $\mathrm{%
Tl_{2}Ba_{2}CuO_{6}}$, and $\mathrm{Tl_{2}Ba_{2}CaCu_{2}O_{8}}$ display
those states, which can be seen in the gray pockets in the corners of the
Brillouin Zone. Another key feature is the region of the Fermi surface
around the $\Gamma $ point, where the gap function approaches zero, a
characteristic absent in other cuprates. Each of the Thallium compounds
contain an electron pocket around the $\Gamma $ point that does not
contribute to superconductivity.

$\mathrm{La_{2}CuO_{4}}$ and $\mathrm{Bi_{2}Sr_{2}CaCu_{2}O_{8}}$ each
exhibit Fermi surfaces that are more diamond--shaped than the other
cuprates, though $\mathrm{La_{2}CuO_{4}}$ has a much simpler Fermi surface.
As doping increases for these materials, the Fermi surface shrinks in a
different manner than the other cuprates. This can lead to a possible
unusual behavior. While this is an active area of research, recent studies
show that significantly overdoped materials of this nature are still
superconducting\cite{Bi_overdoped}.

Experimental work is scarce for $\mathrm{Sr_{2}CuO_{2}Cl_{2}}$ and $\mathrm{%
CaCuO_{2}}$. The latter is the simplest infinite layer cuprate but likely
metastable. Notably the structure of $\mathrm{HgBa_{2}CuO_{4}}$ and $\mathrm{%
CaCuO_{2}}$ are extremely similar as both are tetragonal. Though $\mathrm{%
CaCuO_{2}}$ seems physically not realized in its bulk form, both materials
are expected to give high critical temperatures. $\mathrm{Sr_{2}CuO_{2}Cl_{2}%
}$ has a similar Fermi surface as $\mathrm{HgBa_{2}CuO_{4}}$, though the gap
function is much sharper as the nodes can be seen as taking up less space in
Fig. 1. A gap function with smoother nodes has a smaller kinetic energy and
so is more energetically favorable.

\subsection{b. $\protect\alpha ^{2}F(\protect\omega )$ Spectral Functions}

The Eliashberg spectral functions that we calculate using our LDA+FLEX(RPA)\
method are displayed in Fig. 2, where $\alpha _{x^{2}-y^{2}}^{2}F(\omega )$
represents frequency resolution of the highest eigenvalue $\lambda _{\max }$
corresponding to the $d_{x^{2}-y^{2}}$ symmetry, and $\alpha
_{sf}^{2}F(\omega )$ gives frequency resolution of the mass enhancement
parameter $\lambda _{sf}$. These functions are material--dependent, have
different height, width and shape, and the eigenvalue resolution they
display is unique to each material. However, they also exhibit several
universal features such as the existence of a sharp peak at low frequencies,
followed by a drop--off, which is most pronounced for $\alpha
_{x^{2}-y^{2}}^{2}F(\omega ).$ This drop--off points to a separation of the
energy scales, and justifies the use of McMillan--like $T_{c}$ equation that
relies on the existence of a cutoff frequency for the pairing interaction.

\begin{figure*}[tbp]
\includegraphics[height=0.365\textwidth,width=1.0\textwidth]{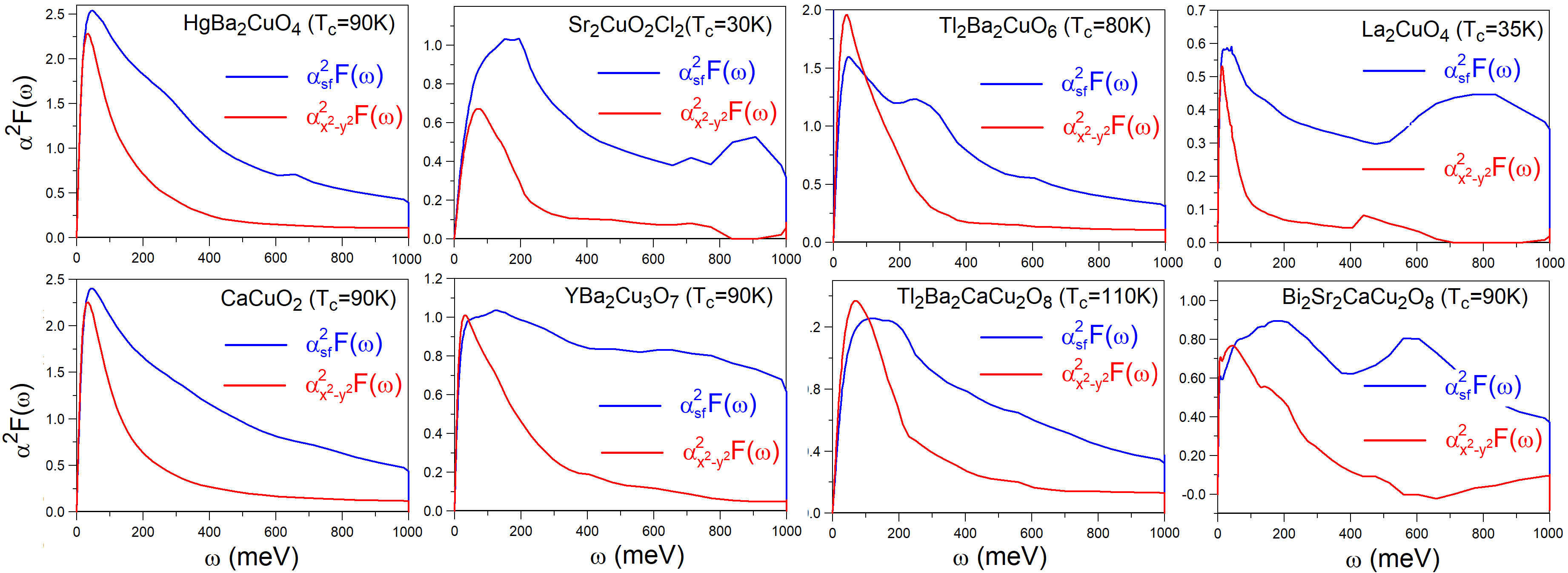}
\caption{Eliashberg spectral functions for 8 different cuprates calculated
using the LDA+FLEX(RPA)\ method. Spectral function $\protect\alpha %
_{x^{2}-y^{2}}^{2}F(\protect\omega )$ (red lines) represents frequency
resolution of the highest eigenvalue $\protect\lambda _{\max }$of the BCS
gap equation corresponding to the $d_{x^{2}-y^{2}}$ symmetry, and $\protect%
\alpha _{sf}^{2}F(\protect\omega )$ (blue lines) gives frequency resolution
of the mass enhancement parameter $\protect\lambda _{sf}$. }
\end{figure*}

The $d$--wave superconducting spectral functions shown by the red lines in
Fig. 2 possess low--frequency peaks most clearly. The spectral functions $%
\alpha _{sf}^{2}F(\omega )$, displayed in blue do not always show this
behavior despite both $\alpha _{sf}^{2}F(\omega )$ and $\alpha
_{x^{2}-y^{2}}^{2}F(\omega )$ originate from the same interaction vertex (%
\ref{INT}), and they are generally expected to exhibit similar shapes. Some
materials have $\alpha _{sf}^{2}F(\omega )$ that somewhat mimic the falloff
in the $d$--wave projection, which for instance, seen in $\mathrm{%
HgBa_{2}CuO_{4}}$ and $\mathrm{CaCuO_{2}}$. Other materials have additional
peaks, beyond the first at low frequency, whose shape does not necessarily
match those of its $d$--wave counterpart.

For all materials, the precise location of the low--frequency peak changes
as a function of $U$. As $U$ gets closer to $U_{SDW}$, the peaks shift to
lower frequencies until they diverge at $\omega =0$ when $U=U_{SDW}$. The
precise values of $U$ that are used to generate the spectral functions in
Fig. 2 are listed in Table 1, column $U_{SC}.$ We select these $U^{\prime }s$
to roughly reproduce the experimental values of $T_{c},$ the procedure to be
described in details in the next section. Here we note that such a choice
generates the low--frequency peaks in $\alpha _{x^{2}-y^{2}}^{2}F(\omega )$
in the region of 40--60 meV. Experimental methods such as neutron scattering%
\cite{SF_review} and ARPES\cite{Vishik_2014} have indicated, respectively,
that there are peaks and kinks found at similar energies for cuprates.
Although the detailed comparison of these values is immature, the agreement
between theory and experiment in the approximate energy scale is a good sign
of the internal consistency in our selected values of $U.$

The effective coupling constant in Eq.(\ref{LAMBDA}) is determined by the
ratio of $\lambda _{max}$ to $\lambda _{sf}+1$. Both $\lambda _{\max }$ and $%
\lambda _{sf}$ being the integrals over the spectral function divided by $%
\omega $, are sensitive to the low--frequency part of the spectrum, that in
turn is most sensitive to $U$. Altering $U$ shifts the location of each
peak, and if these peaks shift together with one function roughly following
the other, then the effective coupling constant $\lambda _{eff}$ evolves
smoothly with $U$ and does not diverge.

We find that cuprates follow this pattern unless $U$ gets very close to its
SDW instability point, $U_{SDW}.$ Fig.3 shows our calculated dependence of
the maximum eigenvalue $\lambda _{\max }$ corresponding to the $%
d_{x^{2}-y^{2}}$ symmetry and the spin fluctuational mass enhancement
parameter $\lambda _{sf},$ as a function of $U$ for the hole doping $\delta
=0.1$ per CuO$_{2}$ plane. The effective coupling constant $\lambda
_{eff}=\lambda _{\max }/(1+\lambda _{sf})$ is shown on the right scale. One
can see that the range of values of $\lambda _{eff}$ is quite modest for a
wide interval of $U$'s as compared to both $\lambda _{\max }$ and $\lambda
_{sf},$ primarily due to the fact that the rise in the eigenvalue of the gap
equation, is compensated by the renormalization effect of the electronic
self--energy. This is, in particular, the case for $\mathrm{HgBa_{2}CuO_{4}}$%
and $\mathrm{CaCuO_{2}}$, each of which have roughly $1eV$ of leeway in the
value of $U$. Interestingly, $\mathrm{YBa_{2}Cu_{3}O_{7}}$ shows similar
behavior: with a very flat $\alpha _{sf}^{2}F(\omega )$, its coupling
constants also keeping pace with each other well as $U$ increases.

Other cuprates have smaller ranges of allowable $U$ interaction strengths
but their $\alpha ^{2}F$'s behave in a similar manner. $\mathrm{%
YBa_{2}Cu_{3}O_{7}}$, $\mathrm{Bi_{2}Sr_{2}CaCu_{2}O_{8}}$ and both
Thallium--based compounds have sharper peaks in $\alpha
_{x^{2}-y^{2}}^{2}F(\omega )$ than their $\alpha _{sf}^{2}F(\omega )$
counterparts. The divergence and shift towards the origin as $U$ increases
affects the respective coupling constants. As the Hubbard term approaches $%
U_{SDW}$, the $\lambda _{max}$ eigenvalue that results from $\alpha
_{x^{2}-y^{2}}^{2}F(\omega )$ increases faster than the mass renormalization 
$\lambda _{sf}+1$, because the low frequency peak is much sharper for $%
\alpha _{x^{2}-y^{2}}^{2}F(\omega )$ and starts to diverge first. This is
seen both in Fig.2 and also in Fig. 3 at the $U$ point where $\lambda _{max}$
meets $\lambda _{sf}$. The impact is seen in how quickly $\lambda _{eff}$
increases in this region. While this effect is not very large for most
materials, it provides a window of opportunity to find the elevated values
of $\lambda _{eff}$ needed to explain the high $T_{c}$'s.

At high energies, though not displayed in Fig. 2, our calculated $\alpha
^{2}F$'s raise rapidly as plasmons begin to possess some spectral weight, a
factor that is normally taken into the $\mu ^{\ast }$ calculations of
McMillan's equation.

\subsection{c. Coupling Constants and the $T_{c}$}

Near the SDW instability, the choice of the Hubbard term $U$ has a
significant impact on the coupling constant and the resulting critical
temperature $T_{c}$. However, methods such as cRPA are not guaranteed to
deliver the precise determination of $U$ as they tend to have much larger
inaccuracies \cite{cRPA_1,cRPA_2}. This prompts us to consider more
empirical pathways.

\begin{figure*}[tbp]
\includegraphics[height=0.466\textwidth,width=1.0\textwidth]{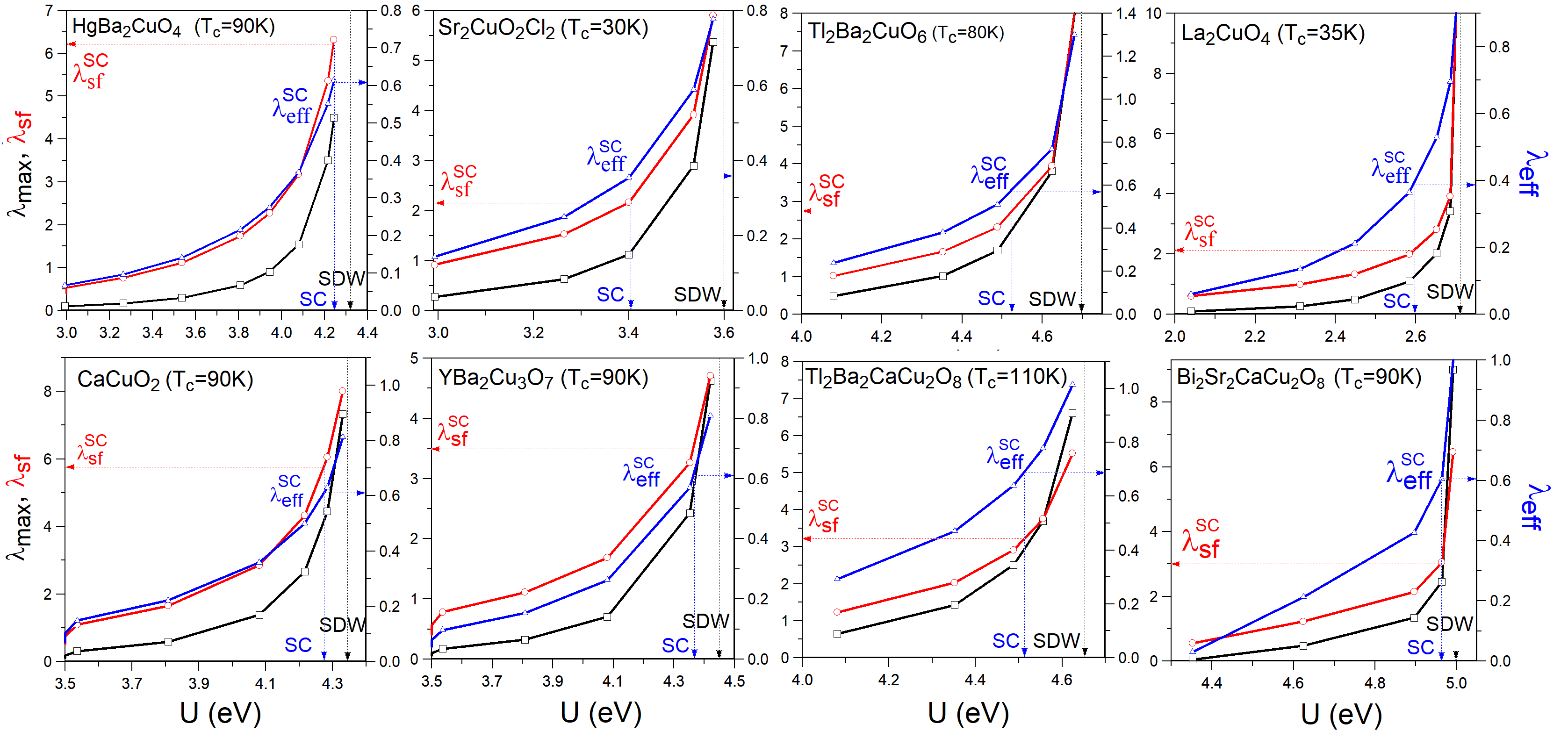}
\caption{{}Calculated using LDA+FLEX(RPA)\ method dependence of the maximum
eigenvalue $\protect\lambda _{\max }$ corresponding to the $d_{x^{2}-y^{2}}$
symmetry and the spin fluctuational mass enhancement parameter $\protect%
\lambda _{sf},$ as a function of the on--site Hubbard interaction $U$ for
d-electrons of Cu in a series of cuprate superconductors for the hole doping 
$\protect\delta =0.1$ per CuO$_{2}$ plane. The effective coupling constant $%
\protect\lambda _{eff}=\protect\lambda _{\max }/(1+\protect\lambda _{sf})$
is shown on the right scale. Sample values are selected with dotted lines
showing how to extrapolate the coupling constants from these plots: $\protect%
\lambda _{\mathrm{eff}}^{SC}$ is selected on the right $y$-axis, giving a
value of $U_{SC}$ on the x-axis, from where the value of $\protect\lambda %
_{sf}^{SC}$ can be determined. Given the nature of the FLEX interaction
terms, the coupling constants diverge in all of these graphs when $U$
approaches the SDW instability point, $U_{SDW}$. }
\end{figure*}

Our calculations of $\lambda _{\mathrm{max}},\lambda _{sf}$ and $\lambda
_{eff}$ as a function of $U$ presented in Fig. 3 allow us to perform
comparisons between theory and experiment. For example, mapping measured
mass enhancement data onto the FLEX--based $\lambda _{sf}$ can determine the
corresponding $U$ for which the experiment can be reproduced, and further $%
\lambda _{eff}$ to see whether one can obtain a reasonable estimate of $%
T_{c} $. In Table 1, we give the compilation of the experimentally extracted 
$\lambda _{sf}$ where the third, fourth, and fifth columns indicate the
values of $\lambda _{sf}$ found from ARPES, quantum oscillations and the
specific heat data\ (figures in brackets indicate the doping at which the
measurement was taken). Although we provide a complete survey of the
experimental results in the next section, here we note the existence of a
rather large spread in the experimentally deduced $\lambda _{sf}$ which
prevents us from concluding on what is the 'right' $U$ to be used by the
theory.

One can take the experimentally known $T_{c}$ and determine such $\lambda
_{eff}$ for which it can be reproduced from Eq.(\ref{TC}). This however
needs an input of the cutoff frequency $\omega _{c}.$ Experimentally, in
cuprates the spin fluctuational frequencies have been seen in the proximity
to 40 meV as peaks in imaginary spin susceptibility accessible via the
numerous neutron scattering experiments\cite{INS}. A famous 40 meV resonance
has been detected in the superconducting state\cite{40meV}. Numerous ARPES
experiments showed kinks in the one--electron spectra at the same energy
range\cite{SF_3}. These kinks are sometimes interpreted as due to the
electron--phonon interactions\cite{Lanzara}, but, unfortunately, the
calculated values of $\lambda _{e-p.}$ are known to be small in the cuprates%
\cite{Savrasov-OKA,Louie}.

Taking $\omega _{c}=40$ meV$,$ and the experimental $T_{c}$ (shown for each
cuprate in the second column of Table 1), we have determined the values of $%
\lambda _{eff}$ by inverting Eq.(\ref{TC}). These values are indicated in
Fig. 3 by labeling $\lambda _{eff}$ with abbreviation 'SC' (superconducting)$%
.$ One can subsequently use this data in Fig. 3 to find the corresponding
values of $U$ and $\lambda _{sf}.$ We indicate the obtained $U$ by arrow
'SC' on the horizontal axis of Fig.3 and the obtained mass enhancement
parameter by $\lambda _{sf}^{SC}.$ We also indicate by arrow on horizontal
axis of Fig. 3 the obtained values of $U_{SDW}$ that correspond to the SDW
divergence. All data deduced from this procedure, $\lambda
_{eff}^{SC},\lambda _{sf}^{SC},\lambda _{\max }^{SC},U_{SC}$ are also listed
in corresponding columns of Table 1.

One can see that in order to match experimental critical temperatures, $%
U_{SC}$ needs to be very close to $U_{SDW}$. In every cuprate displayed, $%
U_{SC}$ is less than $0.2$ eV away from $U_{SDW}$. Different materials
diverge faster or slower than others, giving less or more range of allowable 
$U$ values. $\mathrm{HgBa_{2}CuO_{4}}$ and $\mathrm{YBa_{2}Cu_{3}O_{7}}$
both have this wide range. The materials with steeper dependence of $\lambda 
$ vs. $U$ have much more restricted values of $U$, particularly $\mathrm{%
Bi_{2}Sr_{2}CaCu_{2}O_{8}}$, for which the allowed value of $U$ can be found
with much higher accuracy.

By using the same $\omega _{c}=40$ meV \cite{Neutron_example}, comparisons
of materials with similar critical temperatures can be made. $\mathrm{%
HgBa_{2}CuO_{4}}$, $\mathrm{CaCuO_{2}}$, $\mathrm{YBa_{2}Cu_{3}O_{7}}$, and $%
\mathrm{Bi_{2}Sr_{2}CaCu_{2}O_{8}}$, all have an effective $\lambda $ just
above 0.6. The relatively low choice of $\omega _{c}$ pushes their $U_{SC}$
values very close to $U_{SDW}$ in order to reach the desired transition
temperature.

One can also compare the estimated $\lambda _{sf}^{SC}$ deduced from this
procedure with their measured values. $\mathrm{HgBa_{2}CuO_{4}}$ requires $%
\lambda _{sf}^{SC}=6.2$, which looks above the highest reported measurement,
and so falls out of experimental bounds. Similar cases are seen for other
superconductors, where $\lambda _{sf}$ is seen to be typically overestimated
by this theory although the spread in the experimental values is also large.
It seems likely that the universal choice of $\omega _{c}=40$ meV is not the
best option in trying to find a good match between theory and experiment.

We finally note that as hole doping increases, the Fermi surface shrinks as
electrons are removed from the system, causing the nesting vector to become
smaller. Thus, within our RPA treatment, the gap function gradually begins
to incorporate a $g$--wave solution. This result contradicts with a
well--established experimental result that cuprates exhibit $d$--wave
superconductivity at all dopings, with $T_{c}$ showing a superconducting
dome and decreasing on the overdoped side of the phase diagram. Several
recent studies beyond RPA find the $d$--wave solution\cite%
{two_pairing_domes,FLEX+DMFT}, which reasonably mimics the shape of the
superconducting dome. The presented approach can recover this behavior only
if one allows the dependency of the Hubbard $U$ vs. doping.

Though cRPA values are imprecise, several publications did indeed find that $%
U$\ decreases slightly with doping. A recent work\cite{cRPA_1} reported
computation of doping dependent $U$ using the constrained RPA procedure.
Their reported values of $U\approx $4 eV\ for HgBa$_{2}$CuO$_{4}$ are very
close to the ones needed to produce large $\lambda _{\max .},$ as seen in
our Fig. 3, together with the trend that $U$ decreases with doping a little
bit. In a different work employing dynamical cluster approximation (DCA) 
\cite{DCA-U}, an effective temperature dependent coupling $\bar{U}(T)$ was
introduced to parametrize the DCA\ pairing interaction in terms of the spin
susceptibility. It was extracted between 2 and 4 eV and was shown to exhibit
some reduction upon doping.

\begin{table*}[tbp]
\caption{Compilation of experimental and theoretical results for the
high--temperature superconducting cuprates studied in this work. For each
material, listed are its $T_{c}(K)\ $and mass enhancement parameter deduced
from ARPES, $\protect\lambda _{sf}^{ARPES},$ quantum oscillation, $\protect%
\lambda _{sf}^{dHvA},$ and specific heat, $\protect\lambda _{sf}^{sp.heat}$
experiments. The values in brackets indicate the doping level (number of
holes per one CuO$_{2}$ plane) for which the measurement was taken. Also
listed are the values of Hubbard $U_{SDW}$ $(eV)$ that correspond to the SDW
instability point seen via divergency of the static susceptibility, and the
values of $U_{SC}$ $(eV)$ that are empirically determined from Fig.3 to
reproduce the experimental $T_{c}$ via the BCS $T_{c}$ equation with $%
\protect\omega _{c}=40meV.$\ The theoretical coupling constants $\protect%
\lambda _{eff}^{SC},\protect\lambda _{sf}^{SC},\protect\lambda _{\max }^{SC}$
that correspond to $U_{SC}$ are also given.}%
\begin{tabular}{|c|c|c|c|c|c|c|c|c|c|}
\hline
Compound & $T_{c}^{exp}(K)$ & $\lambda _{sf}^{ARPES}[\delta ]$ & $\lambda
_{sf}^{dHvA}[\delta ]$ & $\lambda _{sf}^{sp.heat}[\delta ]$ & $\lambda
_{eff}^{SC}$ & $\lambda _{sf}^{SC}$ & $\lambda _{max}^{SC}$ & $U_{SC}(eV)$ & 
$U_{SDW}(eV)$ \\ \hline
$\mathrm{HgBa_{2}CuO_{4}}$ & $90$\cite{Hg-Supra} & $0.97[0.13]$\cite%
{Vishik_2014} & 
\begin{tabular}{c}
$3.11[0.096]$\cite{Harrison_2019} \\ 
$2.09[0.1]$\cite{Harrison_2019} \\ 
{$1.59[0.101]$}\cite{Harrison_2019} \\ 
{$1.39[0.115]$}\cite{Harrison_2019} \\ 
{$2.08[0.133]$}\cite{Harrison_2019} \\ 
{$2.42[0.137]$}\cite{Harrison_2019} \\ 
{$3.68[0.152]$}\cite{Harrison_2019} \\ 
{$1.62[0.09]$\cite{Greven_2013}} \\ 
$1.89[0.09]$\cite{Greven_2016}%
\end{tabular}
& $5.42[0.09]$\cite{Klein_2020,Tallon_1994} & $0.61$ & $6.20$ & $4.39$ & $%
4.24$ & $4.32$ \\ \hline
$\mathrm{Sr_{2}CuO_{2}Cl_{2}}$ & $30$\cite{Sr2CuO2} & $1.5[0.0]$\cite%
{Birgeneau_1994} & --- & --- & $0.36$ & $2.22$ & $1.16$ & $3.40$ & $3.60$ \\ 
\hline
$\mathrm{Tl_{2}Ba_{2}CuO_{6}}$ & $80$\cite{Tl-Supra1} & $\sim 4[0.26]$\cite%
{Peets_2007} & 
\begin{tabular}{c}
$1.88[0.297]$\cite{Carrington_2010} \\ 
{$1.94[.270]$}\cite{Carrington_2010} \\ 
{$2.4[0.304]$}\cite{Carrington_2010} \\ 
{$2.42[0.3]$\cite{Wilson_2009}}%
\end{tabular}
& {$1.59[0.33]$\cite{Klein_2020,Tallon_1994}} & $0.57$ & $2.75$ & $2.14$ & $%
4.53$ & $4.70$ \\ \hline
$\mathrm{La_{2}CuO_{4}}$ & $35$\cite{BM} & 
\begin{tabular}{c}
{$2.15[0.095]$\cite{ARPES-MASS2}} \\ 
{$2.85[0.125]$\cite{ARPES-MASS2}} \\ 
{$1.0[0.30]$\cite{Uchida_2003}} \\ 
{$1.1[0.15]$\cite{Uchida_2003}} \\ 
{$2.87[0.063]$\cite{Raffy_2005}}%
\end{tabular}
& {$1.53[0.16]$.\cite{Crooker_2021}} & 
\begin{tabular}{c}
{$0.5[0.33]$\cite{Klein_2020,Hussey_2003}} \\ 
{$2.4[0.22]$\cite{Klein_2019}} \\ 
{$0.6[0.14]$\cite{Klein_2019}} \\ 
{$2.08[0.24]$\cite{Klein_2019}} \\ 
{$2.3[0.25]$\cite{Klein_2019}}%
\end{tabular}
& $0.39$ & $2.10$ & $1.21$ & $2.60$ & $2.71$ \\ \hline
$\mathrm{CaCuO_{2}}$ & $90$\cite{IL-Supra} & --- & --- & --- & $0.61$ & $%
5.57 $ & $4.01$ & $4.27$ & $4.34$ \\ \hline
$\mathrm{YBa_{2}Cu_{3}O_{7}}$ & $90$\cite{123-Supra} & 
\begin{tabular}{c}
{$1.27[0.073]$\cite{Follath_2006}} \\ 
{$1.05[0.095]$\cite{Follath_2006}} \\ 
{$1.03[0.18]$\cite{Follath_2006}} \\ 
{$3.1[0.14-0.18]$\cite{Muller_1988}}%
\end{tabular}
& 
\begin{tabular}{c}
{$0.45[0.1]$\cite{Proust_2008}} \\ 
{$0.57[0.1]$\cite{Taillefer_2007}} \\ 
{$0.24[0.109]$\cite{Klein_2018}}%
\end{tabular}
&  & $0.61$ & $3.50$ & $2.74$ & $4.36$ & $4.45$ \\ \hline
$\mathrm{Tl_{2}Ba_{2}CaCu_{2}O_{8}}$ & $120$\cite{Tl-Supra2} & {$1.8[0.16]$%
\cite{Vishik_2009}} & --- & --- & $0.69$ & $3.20$ & $2.90$ & $4.52$ & $4.66$
\\ \hline
$\mathrm{Bi_{2}Sr_{2}CaCu_{2}O_{8}}$ & $90$\cite{BSCO-Supra} & 
\begin{tabular}{c}
{$0.3[0.23]$\cite{Bi_overdoped}} \\ 
{$1.3[0.106]$\cite{Bi_overdoped}} \\ 
{$0.7[0.16]$\cite{Bi_overdoped}} \\ 
{$0.87[0.111]$\cite{Berger_2006}} \\ 
{$1.0[0.115]$\cite{Berger_2006}} \\ 
{$1.28[0.12]$\cite{Shen_1999}} \\ 
{$1.3[0.135]$\cite{Shen_1999}}%
\end{tabular}
&  &  & $0.61$ & $3.00$ & $2.44$ & $4.96$ & $5.00$ \\ \hline
\end{tabular}%
\end{table*}

\section{IV. Analysis of Experiment}

Numerous studies have been conducted in the past to determine the electronic
mass enhancement parameter $\lambda _{sf}$ in cuprates. Usually, it can be
found from one of three types of experiment: angle resolved photoemission,
quantum oscillation, and specific heat measurements. We focus on seven
cuprates and have extracted the values of $\lambda _{sf}$ from the published
data, displayed in Table 1. This data is compared with theoretical results
found from the LDA+FLEX method.

There is considerable experimental uncertainty in both the value of $\lambda
_{sf}$ and its position on the doping axis of the phase diagram. Doping
levels are often mapped based on $T_{c}$, which introduces imprecision in
nominal doping\cite{Vishik_2014}. Experiments are also quite sensitive to
disorder induced scattering which can lead to additional inaccuracies in $%
\lambda _{sf}.$

There is a variation in experimental conditions that may affect this data as
well. Specific heat measurements take into account all bands, not just those
that contribute to superconductivity, and so can mislead our comparisons
with the values of electronic mass renormalization that we calculate for
correlated electrons only. Quantum oscillation experiments employ very high
magnetic fields. The specific heat data taken at temperatures below $T_{c}$
sometimes also employ magnetic fields to suppress superconducting phase.
These conditions may alter the measured renormalization constants. APRES is
generally a well--regarded method to determine normal--state electronic
structure, but there can still be issues extrapolating $\lambda _{sf}$ since
it is a surface sensitive technique.

In Table 1, the renormalization parameters $\lambda _{sf}$ extracted from
experiment are taken either by direct quote from the publication, or
approximately deduced from the paper's figures. Though straightforward, the
method of extraction depends on the type of experiment and is discussed
below.

\subsection{a. Angle Resolved Photoemission}

ARPES measures the energy band dispersions as a function of a two
dimensional wavevector and neglects the $k_{z}$ dispersion, but for cuprates
with their quasi--2D crystal structures this is usually not a problem. Only
some studies\cite{Uchida_1992} note the non--zero c--axis conductivity,
which, may be an issue for multilayered cuprates. Finite resolution makes
band measurements inexact, and carry possibilities of missing some of the
electronic spectral weight. Other studies note that ARPES shows
significantly more scattering than Angular Magnetoresistance Oscillation
(AMRO) experiments\cite{Peets_2007}, and another study\cite{Harrison_2015}
claims that there is a nonzero density of states in the antiferromagnetic
region that is too small for ARPES to spot.

Extracting data for $\lambda _{sf}^{ARPES}$ that we show in Table 1 means
finding the ratio of effective mass to the calculated band mass, $m^{\ast
}/m_{LDA}$. This is done through the relation $m^{\ast }=\hbar
k_{F}/v_{F}^{\ast }$, where by determining the Fermi momentum and the Fermi
velocity the effective mass is found. Many ARPES studies show energy vs.
momentum plots at different points along the Fermi surface. In this paper,
the $v_{F}^{\ast }$ value is taken at the point at or near the gap node as
this is closest to the Fermi surface. Determining $k_{F}$ at this point was
done by finding the distance from $\Gamma $ to the gap node in the plotted
Fermi surface. Experimental values for the unit cell dimensions are taken
for each material in order to find $k_{F}$. The energy vs. momentum curves
often have significant width, the peak of which is found by mapping the
functions onto a Lorentzian. These peaks are then tracked in order to find $%
v_{F}^{\ast }$. Determining the Lorentzian is not practical without the
original data, and so we visually approximate the peak locations used in the
determination of $v_{F}^{\ast }$. Finally after $v_{F}^{\ast }$ and $k_{F}$
are determined, they are used to calculate $m^{\ast }$.

\begin{figure*}[tbp]
\includegraphics[height=0.5\textwidth,width=1.00\textwidth]{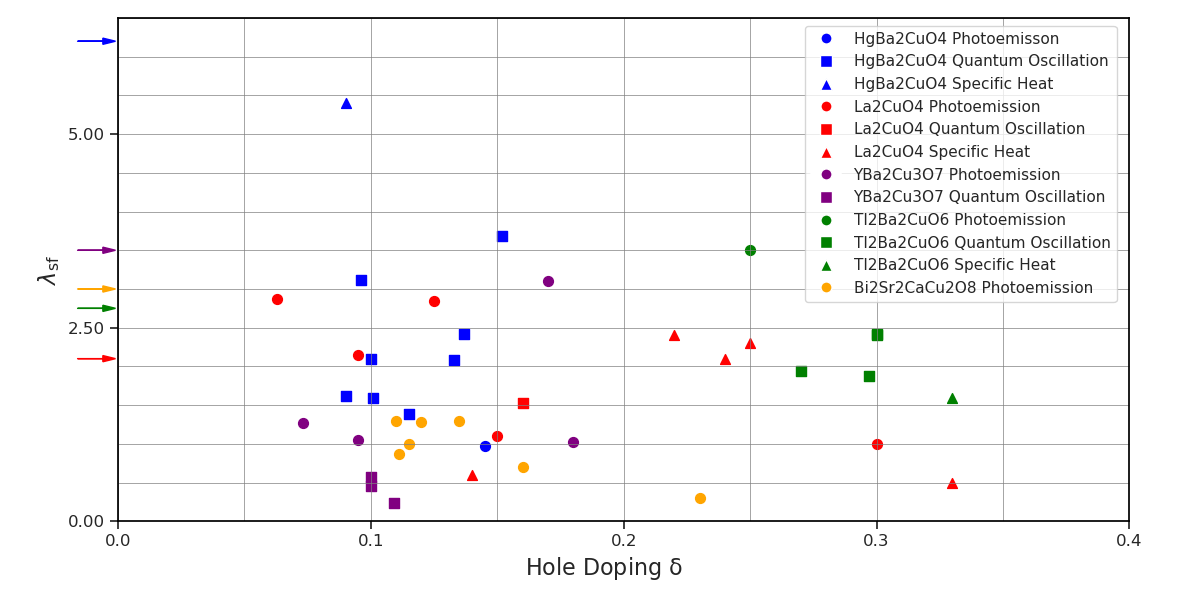}
\caption{Experimental mass enhancement parameter $\protect\lambda _{sf}$
plotted vs doping for several cuprate superconductors studied in this work.
Various colors refer to various materials as listed on this figure. The
shape of the data point denotes the experimental technique from where $%
\protect\lambda _{sf}$ is extracted: ARPES(circles), quantum oscillation
(squares), and specific heat (triangles). Theoretically calculated $\protect%
\lambda _{sf}$ are shown on the vertical axis by arrows with the color
convention as above.}
\end{figure*}

\subsection{b. Quantum Oscillations}

The de Haas--van Alphen (dHvA) effect relies on the electrons oscillating
around the Fermi surface to find values of their effective mass, $m^{\ast }$%
. Unfortunately, often scattering prevents quasiparticles from completing
their orbit, which can throw off measured values. Additionally, one might
expect similar results to be found across different experimental techniques,
but AMRO may detect orbits that are not present in ARPES experiments. This
is potentially due to the Fermi surface reconstruction, the origins of which
are hotly debated\cite{Harrison_2019,Das_2012,Taillefer_2007}. We take the
effective masses directly from figures or values quoted in each publication.
These data are divided by the LDA band masses in order to find $\lambda
_{sf}^{dHvA}$ quoted in Table 1.

\subsection{c. Specific Heat}

Specific--heat measurements determine the mass enhancement by finding the
coefficient $\gamma $ of the electronic specific heat. This is related to
the quasiparticle density of states by $\gamma ^{\ast }/\gamma =N^{\ast
}(0)/N(0)$. The inclusion of additional bands has a possibility of altering
our comparisons with the theory, especially for materials where not all
bands are due to correlated Cu $d$--electrons and contribute to
superconductivity, which, \textit{e.g.}, is the case of $\mathrm{%
YBa_{2}Cu_{3}O_{7}}$. We present $\lambda _{sf}^{sp.heat}$ in Table 1 by
dividing experimental density of states by the LDA determined DOS.

\subsection{d. Discrepancies}

The discrepancies between photoemission, quantum oscillation, and specific
heat measurements can be quite significant. This is seen from our Table 1
and also from Fig. 4 that plots $\lambda _{sf}$ vs. doping, with various
colors referring to various materials and with the shape of the data point
denoting a particular experimental technique: ARPES(circles), quantum
oscillation (squares), and specific heat (triangles). We also show our
calculated $\lambda _{sf}$ on the vertical axis by arrows following the same
color convention. One can generally expect that the mass enhancement is
large in the underdoped, strongly correlated regime, but should diminish
significantly in the overdoped samples. Unfortunately, this tendency is hard
to recognize from the data presented on Fig. 4, although the specific heat
data alone (triangles) do point to this trend.

Some quantum oscillation and specific heat measurements have detected a
second divergence of the mass enhancement in the cuprate phase diagram\cite%
{Greven_2013,Harrison_2019}, near the doping level $0.22$ where the
pseudogap disappears, but this divergence has not been not seen in ARPES. We
have omitted these divergent points from Fig.4.

Another question is whether superconductivity competes with or coexists
alongside surrounding phases, such as the pseudogap, charge density wave
(CDW), and strange metal. This issue remains under debate\cite%
{Zhou_2009,Kaminski_2009,Klein_2018}, and the impact of these phases on the
experimental extraction of $\lambda _{sf}$ is not exactly clear. Also, our
work does not address any potential ground state competition between these
phases and superconductivity.

\section{V. Conclusion}

In conclusion, we have studied a series of cuprate superconductors using our
recently developed LDA+FLEX(RPA) method that accounts for the electronic
self--energy of the correlated electrons using a summation of the
particle--hole bubble and ladder diagrams. Based on this procedure,
superconducting energy gaps, Eliashberg spectral functions $\alpha
^{2}F(\omega ),$ and spin fluctuational coupling constants have been
calculated using realistic energy bands, wave functions, and the Fermi
surfaces of the cuprates. The Hubbard interaction $U$ is the only parameter
in this method.

The predicted $d_{x^{2}-y^{2}}$ pairing symmetry is a robust feature of our
calculation together with the universal shape of the $\alpha ^{2}F(\omega )$
exhibiting a strong peak at the low--frequency paramagnon region, and its
rapid decay towards higher frequencies justifying applicability of the
BCS--type approximation. However, a strong dependence on$\ U$ of
superconducting coupling constant $\lambda _{\max }$ that comes out as the
maximum eigenvalue of the BCS equation, and the same behavior for the
calculated quasiparticle mass enhancement $m^{\ast }/m_{LDA}=1+\lambda _{sf}$
makes the effective spin fluctuational coupling constant $\lambda
_{eff}=\lambda _{\max }/$($1+\lambda _{sf})$ small unless $U$ is tuned to be
within a few per cent of the SDW instability point that occurs for each
material at some $U_{SDW}.$ This behavior highlighted challenges in building
first--principle theories of high--temperature superconductivity but
provided a window of opportunity to find the elevated values of $\lambda
_{eff}$ needed to explain the high $T_{c}$'s.

We have also analyzed a wealth of published mass enhancement data that can
be extracted from ARPES, quantum oscillation and specific heat experiments.
This data offered additional comparisons with the values of $\lambda _{sf}$
that we calculated in the hope to find further constrains for the material
specific $U^{\prime }s.$ Unfortunately, the spread in the experimentally
deduced values of $\lambda _{sf}$ did not offer further insights.

At the end, we hope that gaining further experience with applications of
this method and its extensions to other unconventional superconductors will
ultimately allow us to reach a more quantitative understanding of
unconventional superconductivity in cuprates and other systems.

\end{document}